\documentclass[12pt,preprint]{aastex}
\usepackage{natbib}
\usepackage{graphicx}
\usepackage{ulem}

\shorttitle{Initiation of flare-CME event on 03-November-2010}
\shortauthors{Mulay et al.}

\begin{document}

\title{Initiation of CME event observed on November 3, 2010: Multi-wavelength Perspective}
\author{Sargam Mulay, Srividya Subramanian, Durgesh Tripathi}
\affil{Inter-University Centre for Astronomy and Astrophysics, Post Bag - 4, Ganeshkhind, Pune 411007, India}

\author{Hiroaki Isobe}
\affil{Center for the Promotion of Interdisciplinary Education and Research, Kyoto University, Kyoto 606-8501, Japan}

\author{Lindsay Glesener}
\affil{Space Sciences Laboratory, University of California at Berkeley, 7 Gauss Way, Berkeley, CA 94720, USA}

\begin{abstract}

One of the major unsolved problems in Solar Physics is that of CME initiation. In this paper, we have studied the initiation of a flare associated CME which occurred on 2010 November 03 using multi-wavelength observations recorded by Atmospheric Imaging Assembly (AIA) on board Solar Dynamics Observatory (SDO) and Reuven Ramaty High Energy Solar Spectroscopic Imager (RHESSI). We report an observation of an inflow 
structure initially in 304~{\AA} and in 1600~{\AA} images, a few seconds later. This inflow strucure was detected as one of the legs of the CME. We also observed a non-thermal compact source concurrent and near co-spatial with the brightening and movement of the inflow structure. The appearance of this compact non-thermal source, brightening and movement of the inflow structure and the subsequent outward movement of the CME structure in the corona led us to conclude that the CME initiation was caused by magnetic reconnection.

\end{abstract}


\section{Introduction} \label{intro}

Coronal Mass Ejections (CMEs) are the most energetic Solar Events. These are considered to be the direct consequence of
dynamics of the solar atmosphere. CMEs are one of the fundamental processes responsible for transfer of mass and energy from 
the Sun into the heliosphere. They represent potential threats to the space weather and geo-space climate. In spite of considerable 
advancement in the quality of the observational data and numerical simulations, the initiation of CMEs is still a matter of strong debate. 
For reviews see e.g., \citet{schwenn2006,gopal2006,wh2012}.

Observationally, the evolution of CMEs have been found to have mainly three different stages, namely - initiation, acceleration and 
propagation. The initiation phase, also called as the slow rise phase, often found to be associated with the pre-flare phase  of the 
associated Soft X-ray flaring emission \citep[e.g.,][]{chifor_2006}. While the acceleration phase of the CME often corresponds to 
magneto-hydrodynamic instability and main phase or impulsive phase of the flare and the propagation phase of the CME corresponds 
to the decay phase of the flare \citep[see e.g.][ and references therein]{Zhang2006,chifor_2006,chifor_2007,Tripathi2013}. 

\begin{figure*}
\centering
\includegraphics[width=0.8\textwidth]{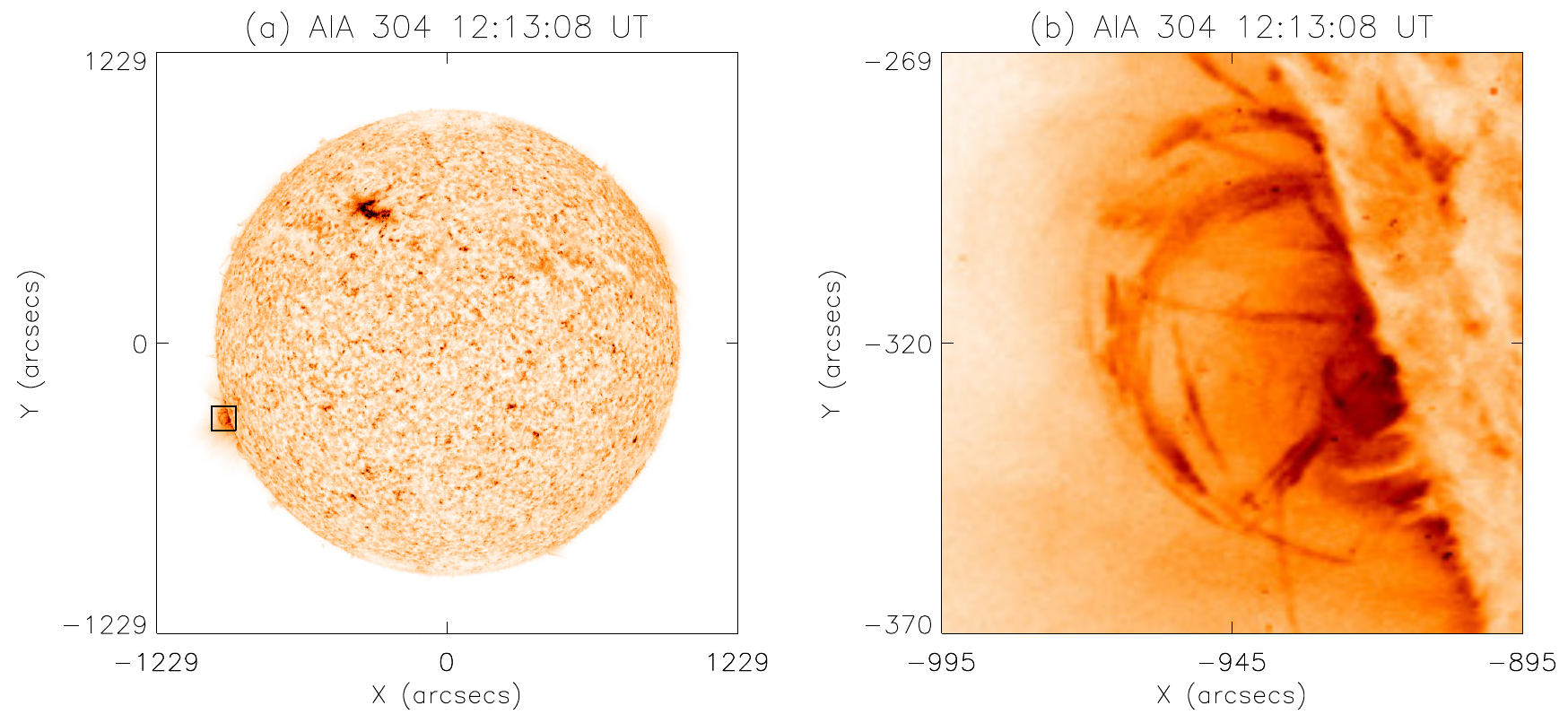}
\caption{Left panel: Full disk image of the Sun at 304~{\AA}. The over-plotted box shows the region where the CME eruption took place. 
Right Panel: 304~{\AA} image corresponding to the over-plotted box shown in the left panel.}
\label{fig1}
\end{figure*}

Magnetic reconnection \citep{Parker1963,Petschek1964} has been considered to play an important role in the initiation of almost all the 
dynamic and eruptive solar events such CMEs \citep[][]{SteM:2005,chifor_2006,chifor_2007}, flares  
\citep[see e.g.,][]{Sweet1958,Shibata1996,Priest2002,Tripathi2004,su_2013}, partial eruption of prominences 
 \citep[][]{GibF:2006,Tripathi2006a,Tripathi2006b,Tripathi2009a,Tripathi2013} coronal jets \citep{Shibata1992,Shibata1994,Shibata2008, chifor_2008},  
 etc observed in the solar atmosphere. There have been numerous indirect evidences for the reconnection phenomena such as reconnection inflow 
 \citep{Yokoyama2001,Savage2012,Takasao2012} and reconnection outflow \citep{McKenzie1999, McKenzie2001,Tripathi2006b,Tripathi2007,Savage2010}, 
 non-thermal hard X-ray sources associated with initiation of CMEs \cite{chifor_2006}. However, these different signatures have been observed individually 
 in separate dynamic events. Coronal mass ejections are almost always associated with solar flares, for which reconnection is a very important concept \citep{Benz2008}.

\begin{figure}
\centering
\includegraphics[width=0.8\textwidth]{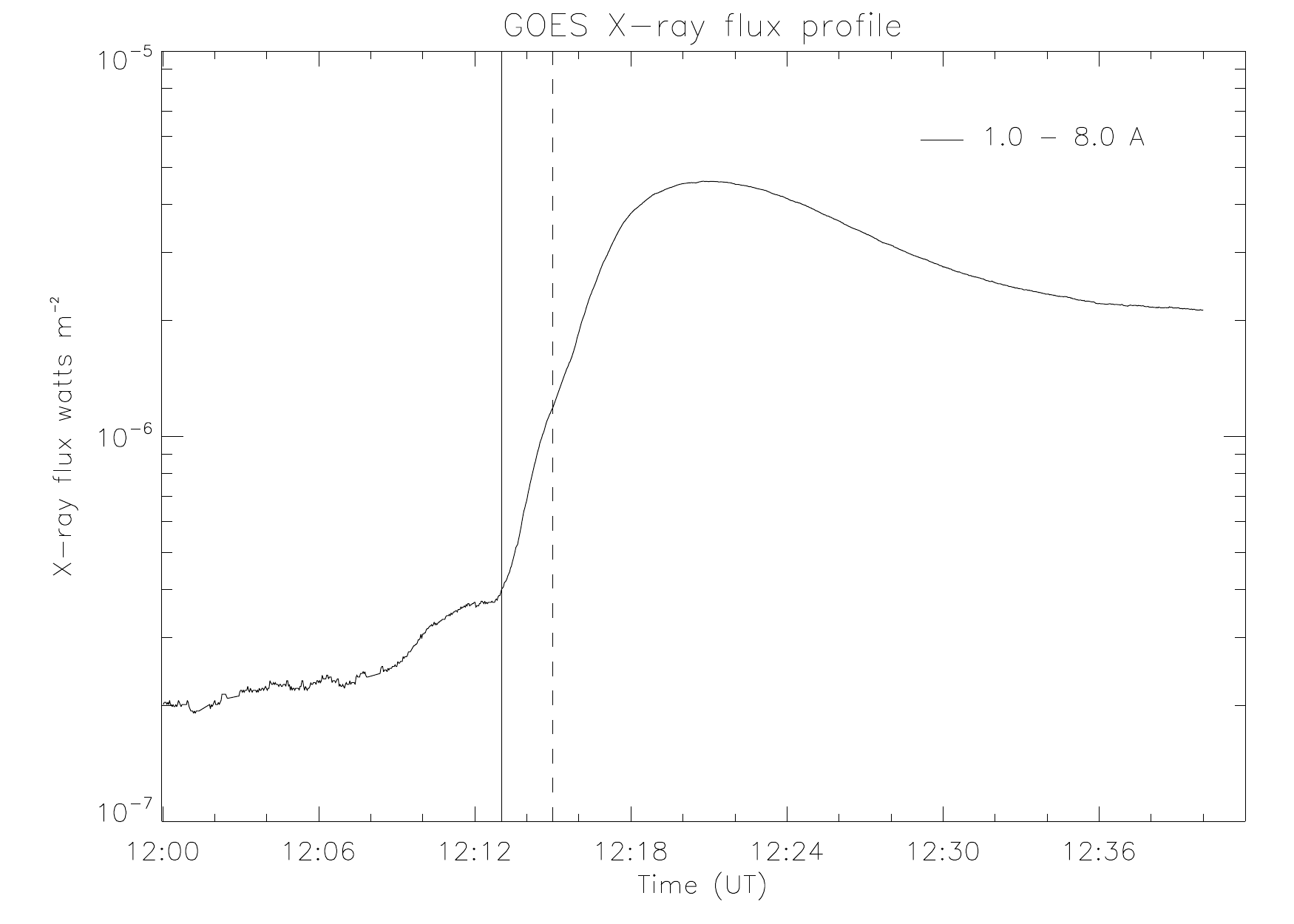}
\caption{Integrated Soft X-ray flux (1-8~{\AA}) observed by GOES 14 satellite on November 03, 2010. The continuous vertical line shows the beginning of 
the event at about 12:13:08 UT and the dotted vertical line marks the end at about 12:15:20 UT, when the 304~\AA~inflow loop disappears completely.}
\label{fig2}
\end{figure}

A flare associated CME was observed on 2010 November 03 originating from the partially occulted active region \textit{AR~11121}. This event 
was well observed by Atmospheric Imaging Assembly (AIA) on board Solar Dynamics Observatory (SDO), Reuven Ramaty High Energy Solar 
Spectroscopic Imager (RHESSI) and radio instruments.  Number of papers have been published addressing  various issues associated 
with flares and CMEs. For example, the evolution phase of the CME and associated flare is studied by 
\cite{Reeves2011,Savage2012,Kumar2013,Foullon2011,Cheng2011,Hannah2013, Glesener2013} including the relationship between flares and CMEs. 
The associated decimetric/metric type II radio burst is analysed by \cite{Bain2012,Zimovets2012}. Moreover, this event showed the first evidence of magnetic Kelvin-Helmholtz instability in the solar corona which were reported by \cite{Foullon2011, Foullon2013}. However, none of the above papers studied the near-limb region of the event nor considered the initiation phase of the CME. 

The main aim of this paper is to study the initiation of this flare-CME event by making use of the multi-wavelength observations recorded by AIA/SDO and RHESSI. By combining these observations, we find that magnetic reconnection is the cause for the initiation of this CME event. AIA images taken at 304~{\AA} and 1600~{\AA} show signature of inflow structure consistent with the time and location of Hard X-ray emission imaged by RHESSI. Rest of the paper is structured as follows. In section~\ref{obs} we report the observations. In section~\ref{Analysis_Result} we describe our data analysis and results followed by the discussion and conclusion in section~\ref{sum}.


\section{Observations} \label{obs}

\begin{figure*}
\centering
\includegraphics[width=0.95\textwidth]{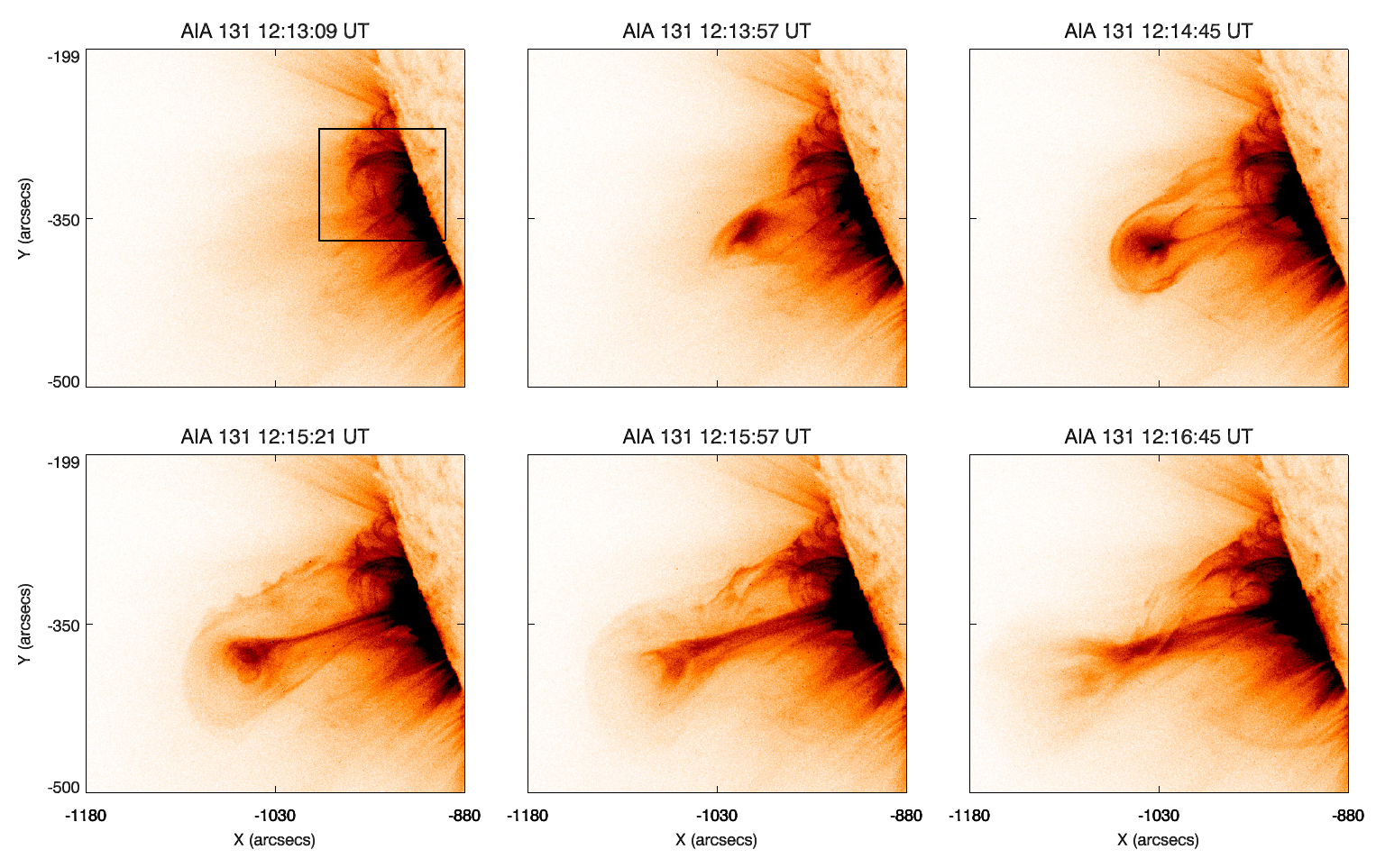}
\caption{Sequence of images taken using 131~{\AA} filter showing the complete evolution of the CME event. The box in the top left panel
shows the field of view which is analysed in detail.}
\label{image131}
\end{figure*}
\begin{figure*}
\centering
\includegraphics[width=0.95\textwidth]{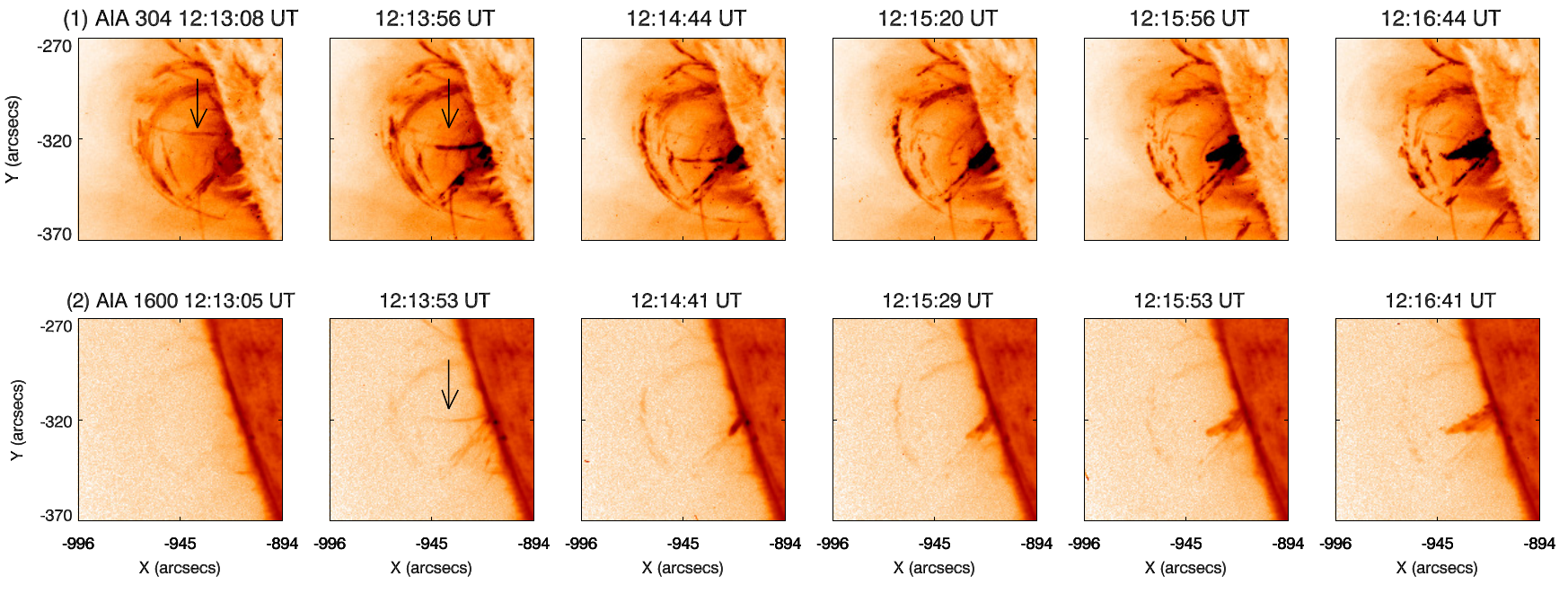}
\caption{Sequence of images during the impulsive acceleration phase of the CME, 
at 304~{\AA} (First row) and 1600~{\AA} (second row). The FOV of these images corresponds
to the box shown in Fig~\ref{image131}. }
\label{fig3}
\end{figure*}

In the current analysis, we have focussed on AIA and RHESSI data obtained during the early rise phase of the CME initiation. The Atmospheric 
Imaging Assembly \citep[AIA; ][]{Lemen2012} onboard Solar Dynamic Observatory (SDO) provides continuous full disk observations of the solar 
atmosphere in 10 different UV/EUV passbands with a high spatial resolution (pixel size of 0.6$\arcsec$), a temporal resolution of 12 seconds 
and a field of view covering up to 1.3 solar radii. we have concentrated only on three AIA channels, namely 1600~{\AA} dominated by \ion{C}{4} 
emission (transition region temperature and the continuum emission from the photosphere), 304~{\AA} dominated by chromospheric \ion{He}{2} at 
quiet conditions and 131~{\AA} dominated by coronal \ion{Fe}{8}, \ion{Fe}{20}, \ion{Fe}{23}. The readers are referred to \cite{O'Dwyer2010}, for detailed description of EUV filters in AIA and their temperature response. The observations recorded using other wavelength channels have been extensively analysed by previous authors. The features studied in the current paper have not been observed in other wavelength channels. All the data are reduced and prepared using the standard procedures available in the solarsoft libraries.

The RHESSI mission \citep[][]{Lin2002} provides high-resolution spectra and images in hard X-rays, from 3 keV to 17 MeV. It has nine sub-collimators, all with different spatial resolutions and each sub-collimator does not modulate flux from sources larger than its resolution \citep{Hurford2002}. In this work, we have used only sub-collimators 3--5 which have resolutions from 7 to 20 arcseconds.

On 03 November 2010, AIA observed an event originating from the East limb at around 12:13~UT. The source region was located 5-7 degree behind the optical limb \cite{Glesener2013}. The source region appeared on the visible disk on 05 November 2010 and was named as \textit{AR11121}. The left panel of Figure~\ref{fig1} displays the full disk image of the Sun taken using 304~{\AA} filter. The small over-plotted box on the East limb represents the 100$\arcsec$ X 100$\arcsec$ area of interest.The right panel shows the close-up of the region shown via box in the left panel. Figure~\ref{fig2} shows three 
seconds averaged soft X-ray flux of the associated C4.9 class flare between 1{--}8~{\AA} obtained using GOES 14 satellite. The profile suggest that the flare started at around 12:07 UT with a peak at around 12:21 UT. A pre-flare bump is also noticeable before the main phase of the flare. 

Figure~\ref{image131} displays the complete evolution of the flare-CME event. A bright loop like structure emanates from the source region at 12:13:09~UT, which later became a CME. With passing time, the structure evolves and shows a fully formed flux rope with a current sheet underneath and overlying loop structure. The over plotted box shows the region which we analyse in detail. 
\section{Data Analysis and Results} \label{Analysis_Result}

Figure~\ref{fig3} show observations of the source region taken with 304~{\AA} (top row) and 1600~{\AA} (bottom row) filters between 12:13 to 12:17~UT, which corresponds to the rise phase during the CME initiation and the impulsive phase of the associated flare (see Fig.~\ref{fig2}). The field of view of these images corresponds to the boxes shown in Fig~\ref{fig1} and Fig~\ref{image131}. The images in both the wavelengths show loop like structures and other filamentary structures. A close analysis of images taken in 304~{\AA} (top row) reveals a pre-existing vertical structure (with respect to the limb; marked with arrows in the first and second panels of Fig.~\ref{fig3}), that moves southward as seen in the subsequent images. We refer to this structure as "inflow structure" hereafter. The vertical inflow structure first brightens up in 304~{\AA} image at 12:12:56 UT. With passing time, as it moves southwards, it becomes brighter and finally disappears at 12:15:20 UT.  By tracking a point on the inflow 
structure at different instances of time until it disappears, we determined the speed which was 105~km~s$^{-1}$. 

\begin{figure*}[!ht]
\centering
\includegraphics[width=1\textwidth]{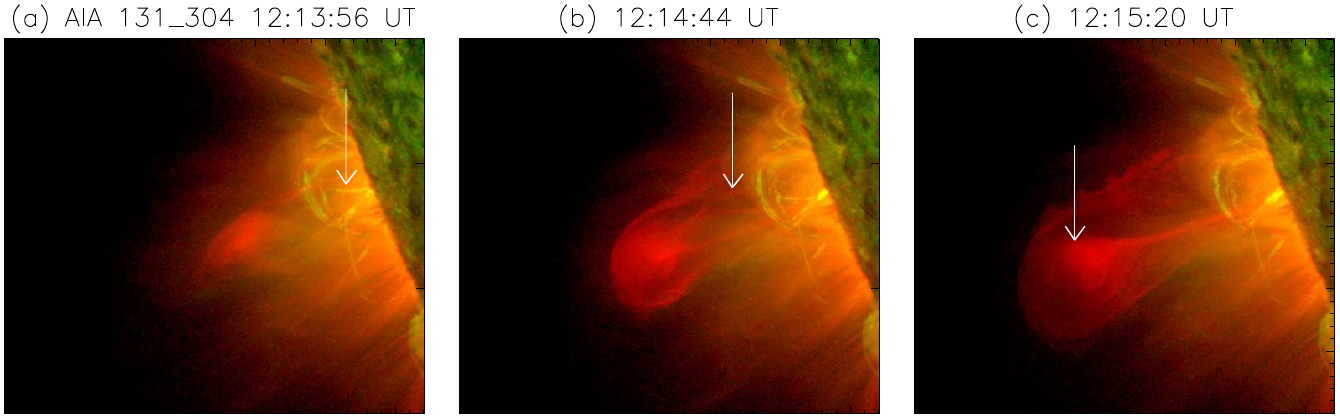}
\caption{Composite images of observations recorded in 304~{\AA} (green) and 131~{\AA} (red). The arrow in the left image shows the inflow 
structure seen in 304~{\AA} and that in the middle image shows the connection between the inflow structure and the northern leg of the flux rope. 
The arrow in the right image shows the fully formed flux rope structure.} \label{fig4}
\end{figure*}

This inflow structure remains invisible in the 1600~{\AA} images until the structure starts moving in 304~{\AA} images. Once the inflow structure starts to move southwards and gets brightened in 304~{\AA}, it becomes visible in the 1600~{\AA} images at about 12:13:28 UT (structure is shown by an arrow in the second panel of the bottom row in Fig~\ref{fig3}). This structure is, however, seen until 12:14:17 UT in 1600~{\AA} images and disappears completely, which coincides with the fading of this structure in 304~{\AA} images. A jet like outward moving structure, similar to that is seen in 304~{\AA} images, is also observed. Absence of this structure in 1700~{\AA} images, confirms that the 1600~{\AA} emission comes mainly from the \ion{C}{4}, i.e. the emitted plasma is at the transition region temperature. 

Figure~\ref{fig4} shows three snapshots of composite images near simultaneously recorded using 304~{\AA} (green) and 131~{\AA} (red). The arrow in the left panel image locates the inflow structure seen in 304~{\AA} images (discussed earlier). The middle panel shows the inflow structure seen in 304~{\AA} is connected to the northern leg of the blob seen in  131~{\AA}, as marked by an arrow. The arrow in the right panel shows the fully formed flux rope like structure. The associated animation \textit{composite\_movie\_131\_304.mpg} reveals the evolution of the CME seen in 131~{\AA} and the inflow structure seen in 304~{\AA}. The composite movie shows a large scale faint disturbance which occurs in close association with the first brightening of the inflow structure in 304~{\AA} images. A diffuse blob-like structure appears and starts to move outwards in concurrence with the inflow structure sweeping in towards the south. The composite images taken at 12:14:20~UT shows a blob like structure with two legs 
emanating from the location where the inflow structure merge with another pre-existing structure. By 12:14:44 UT, the blob like structure takes a shape of a fully formed flux-rope with an overlying loops. We interpret this inflow structure to be reconnection inflow, which moves southward to reconnection location at which it merges with other magnetic structure and disappears.

\begin{figure}[!ht]
\centering
\includegraphics[width=1\textwidth]{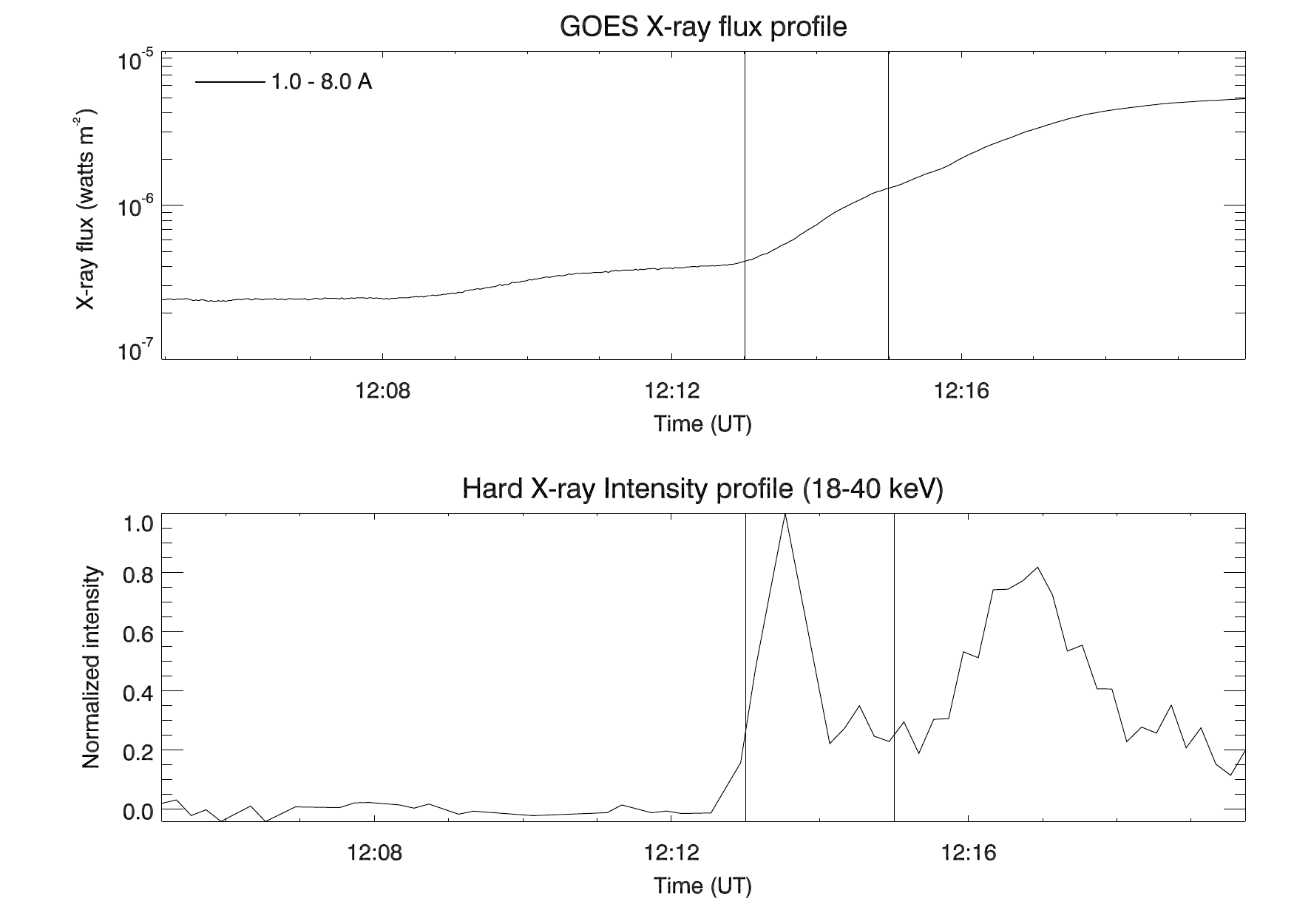}
\caption{Top panel: GOES Soft X-ray profile integrated between 1-8~{\AA}. Bottom panel: RHESSI light curve between 18-40 keV. The two vertical lines mark the time interval
which corresponds to the brightening, movement and disappearance of the inflow structure
seen in 304~{\AA}.} 
\label{goes-rhessi}
\end{figure}
\begin{figure*}[!ht]
\centering
\includegraphics[width=1.0\textwidth]{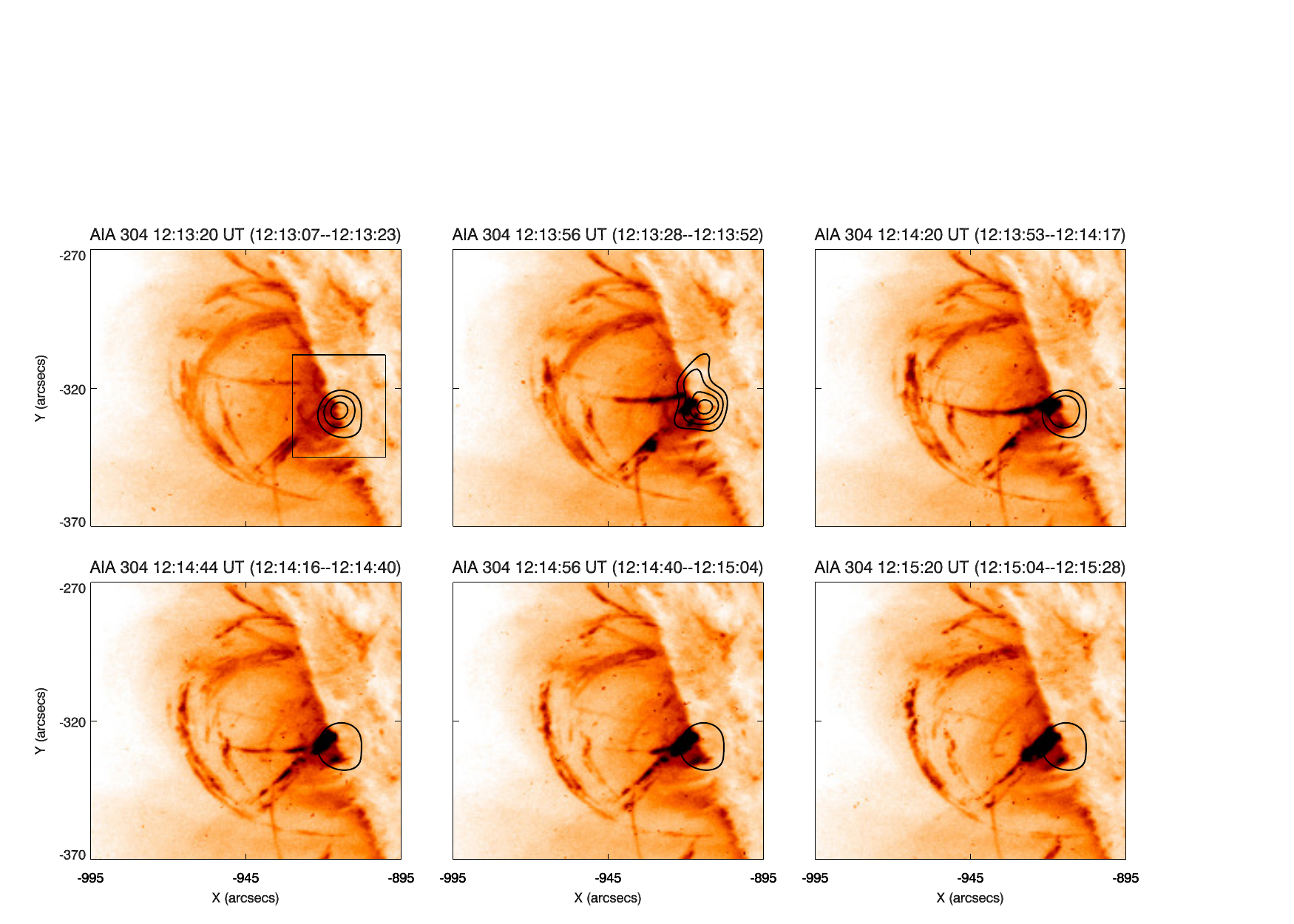}
\caption{Sequence of images observed during the rise phase of CME initiation at 304~{\AA} (color inverted), with overplotted RHESSI contours in 18-40 keV energy band. The contours in each frame corresponds to the ratio of the hard X-ray flux observed in that particular frame with respect to the peak flux observed in the whole event (Figure~\ref{goes-rhessi}:bottom panel). The top-middle panel corresponds to the peak of the hard X-ray light curve. The contours are 30\%, 50\%, 70\%, and 90\% of the maximum flux in the top center image. The respective RHESSI images are created by combining the fine subcollimators (3, 4 and 5) with CLEAN algorithm over the time given in the bracket. The box overplotted in the top-left panel shows the field of view that is used to generate the time profile of the near limb compact hard X-ray source.} \label{aia_rhessi}
\end{figure*}

Distinct Hard X-ray (HXR) emission regions were observed by RHESSI \citep[see Figure 3 in ][]{Glesener2013}. Using 18-40 keV RHESSI observations with two step CLEAN method \citep{Krucker2011}, they observed two distinct HXR sources: a compact source near the limb and a coronal extended diffuse source. The later was co-spatial and co-temporal to the main phase of the flares and associated erupting CME as observed in AIA 131~{\AA} images. 
However, the compact source appeared about 30 seconds prior to the coronal extended diffuse source. The delay could not be explained by the difference in electron densities due to much shorter time for thermalisation of the accelerated electrons.
An explanation could be there were two different reconnection events, one occurring closer to the limb and other at higher altitude in the corona. The start timing and location of the compact source near the limb matches with the brightening and movement of the inflow structure observed in 304~{\AA} 
images shown in Fig.~\ref{fig3}. In this paper we have, therefore, focused on this particular source.

Figure~\ref{goes-rhessi} shows GOES Soft X-ray flux profile integrated between 1-8~{\AA} (top panel) and Hard X-ray (HXR) light curve integrated between 18-40 keV (bottom panel). Right at the beginning of the impulsive phase of the flare, an enhancement in HXR source is seen. First appearance of this HXR emission is at 12:12:56 UT. This also marks the onset of the rise phase of CME initiation and the brightening and movement of 304~{\AA} inflow structure. Moreover, the HXR light curve shows two peaks (bottom panel of Fig.~\ref{goes-rhessi}). The first peak corresponds to the very early times of the impulsive phase of the flare and the intensity is about 20\% larger than the second peak. The two vertical lines mark the time interval which corresponds to the brightening, movement and disappearance of the inflow structure seen in 304~{\AA}. In order to probe the spatial relation between location of the near limb compact HXR source and the inflow structure observed in 304~{\AA} images, we have created HXR images 
with CLEAN algorithm \citep{Hurford2002}. Figure~\ref{aia_rhessi} shows the sequence of (reverse colored) images observed during the course of this event in 304~{\AA} over-plotted with RHESSI contours. The contours in each frame correspond to the ratios of the HXR fluxes observed in that particular frame with respect to the peak flux. The top-middle panel corresponds to the peak timing of the HXR light curve. Hence, it shows a range of contours corresponding to 30\%, 50\%, 70\% and 90\% of the peak flux, respectively. The box over-plotted in the top-left panel shows the field of view that is used to generate the light curve of the near limb compact hard X-ray source shown in Figure~\ref{goes-rhessi}. 
As can be seen from the Figures~\ref{goes-rhessi} and \ref{aia_rhessi}, the HXR emission in 18-40 keV is co-temporal with the inflow motion and co-spatial with the base of the inflow structure, suggesting that this inflow is indeed the reconnection inflow that energizes the flare.
One puzzling point in this interpretation is the brightening of the inflow seen in 304~{\AA} Figure~\ref{aia_rhessi}, because the inflow should be pre-heated plasma. One possibility is that the inflow is heated by the leakage of the heat from the reconnection region by thermal conduction, as suggested by \cite{Yokoyama1997} and \cite{Chen1999}.

\begin{figure}[!ht]
\centering
\includegraphics[width=0.6\textwidth]{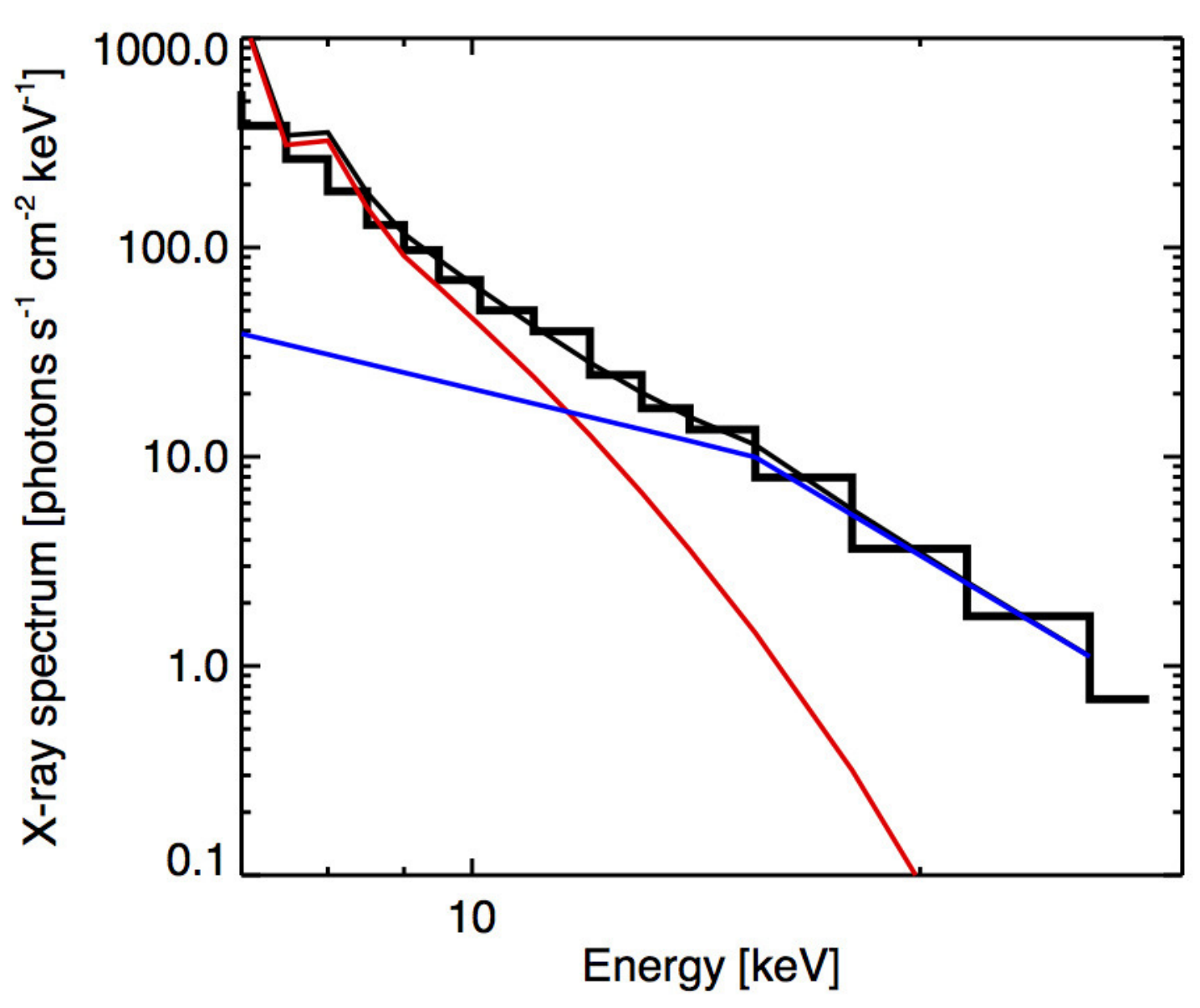}
\caption{Hard X-ray spectrum averaged over 12:12:56 - 12:13:50 UT shown in black. The observations are fitted 
with a thermal distribution in red (T= 21.7 MK, EM = 6.5 $\times$ 10$^{46}$ cm$^{-3}$) and a broken power law in blue 
(power law index is 4.2).}
\label{rhessi-spec}
\end{figure}

We have also performed the spectral analysis of the near-limb HXR source during the time interval 12:12:56 - 12:13:50 UT. The images were individually checked to ensure that Clean components were not found high in the corona, to verify that the images are not contaminated by flux from the high-coronal source. A photon spectrum was fit to the fluxes of these Clean images for this time interval and is shown in Figure \ref{rhessi-spec}; fit components included a thermal distribution (T= 21.7 MK) and a broken power-law with spectral index 4.2. The existence of a power law component above 20 keV confirms a population of flare-accelerated electrons in the near-limb source. This analysis suggests that non-thermal nature of the source, demonstrating magnetic reconnection processes in action, which in turn leads to the movement of the inflow structure seen in 304~{\AA} observations and initiation of the CME event.

\begin{figure*}[!ht]
\centering
\includegraphics[trim = 0.5cm 8cm 2cm 3.5cm, width=0.9\textwidth]{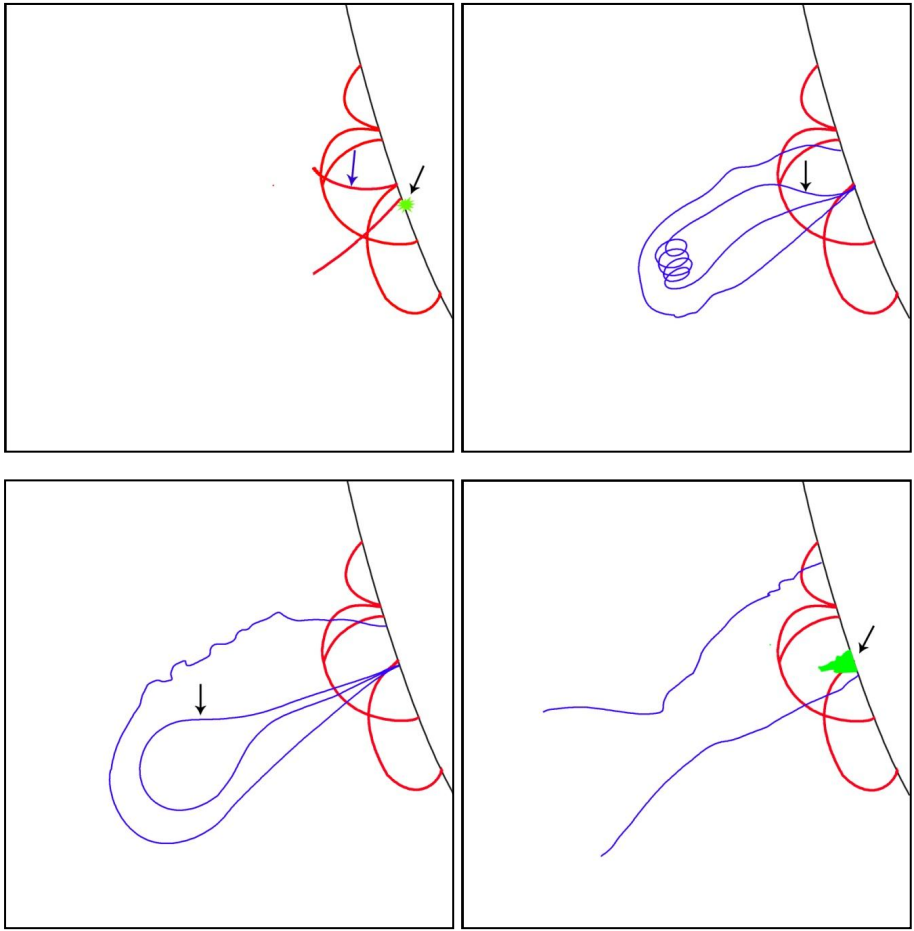}
\caption{Schematic depiction of the complete evolution of flare-CME event observed over the east limb. 
Top left panel: scenario before the event and onset of HXR source (green in color marked by an arrow) over the 
limb. The structure marked with the blue arrow is the inflow structure. Top right panel: inflow structure connected 
to the northern leg of the blob during the rise phase of CME initiation. Bottom left panel: fully formed flux rope 
structure with current sheet underneath and overlying loop structure, Bottom right panel: A jet (green in color) 
like outward moving structure similar to that seen in 304~{\AA} images.}
\label{fig9}
\end{figure*}

Figure~\ref{fig9} depicts the complete evolution of initiation phase of the event. The top left panel shows 
the structures in the source region (shown in red) including the inflow structure (marked with a blue arrow) seen 
in 304~{\AA} image before the CME eruption. The structure depicted by blue line indicates flux rope structure 
as observed in 131~{\AA}. The inflow structure in the top left panel brightens up with the occurrence of non-thermal
bright compact HXR source (18-40 keV) shown in green and marked with a black arrow, observed at 12:12:56~UT. 
Top right panel shows the inflow structure as it appears in 304~{\AA} and is connected 
to the northern leg of the blob seen in 131~{\AA}, as marked by an arrow. The fully formed flux rope structure 
shown in bottom left panel and marked with an arrow later erupts as a CME. The Bottom right panel shows the 
evolution of jet material (shown in green) that appears concurrent with the pinching of the inflow structure. The 
appearance of this compact non-thermal source, brightening and movement of the inflow structure and the 
subsequent outward movement of the CME structure in the corona led us to conclude that the magnetic 
reconnection is the cause for the CME initiation. Based on the location of the HXR source followed by moving 
inflow structure and subsequent UV brightening associated with jet, we infer that the reconnection region is located 
at even lower altitude than that was reported by \cite{Savage2012}.

\section{Summary and Conclusions} \label{sum}
The problem of CME initiation is one of the major unsolved problems in Solar Physics. This has profound impact on space weather forecasting. In this paper, we have probed the initiation phase of a well observed flare associated CME event which occurred on November 03, 2010 using multi-wavelength observations recorded using AIA onboard SDO and RHESSI. The later evolution phase of the CME and associated flare event has been studied in detail by various authors \citep[e.g.,][]{Reeves2011, Savage2012, Kumar2013, Foullon2011,Cheng2011,Bain2012, Zimovets2012, Hannah2013, Glesener2013}. However, no one has studied the initiation phase of the CME, which is the aim of this paper. 

The results are summarised below:
\begin{itemize}

\item An inflow structure was observed in 304~{\AA} images. The same inflow structure appeared after about few seconds in 1600~{\AA} images, 
suggesting heating of the inflow structure, which is also demonstrated by the increase in the brightening of the inflow structure in 304~{\AA} observations.

\item A bubble like structure, which later became a flux rope like structure and erupted as CME, started to rise in 131~{\AA} images exactly at the time when the inflow structure was first seen.

\item A compact HXR source was seen over the limb at about 12:12:56 UT, exactly at the time of the first brightening in the inflow structure followed by its movement. The start of the movement of the inflow structure and appearance of the HXR compact source matched with the time of the rising blob which later manifested as a CME. The light curve of this source unveils that the HXR fluxes increase, until it peaks at about 12:13:30 UT and then fades to a minimum by 12:14:20 UT. The spectral analysis of the source suggests the non-thermal nature of the source, demonstrating the magnetic reconnection processes to be responsible for the HXR emission. 

\end{itemize}

These observations suggest that the initiation of the CME is triggered by magnetic reconnection taking place in the source region. This is demonstrated by brightening of the inflow structures co-temporal with the non-thermal compact HXR source in the source region and outward movement of the blob-like structure which later manifests itself as a CME.

A plausible interpretation could be the following. As it is revealed by the images obtained at 131~{\AA}, there is a blob like structure pre-existing in the corona with the inflow structure 304~{\AA} as one of the legs. This structure is in stable configuration under the Lorentz force balance. Due to the evolution of the magnetic field in the source region, magnetic reconnection takes place producing the compact non-thermal HXR. This could also be produced by magnetic reconnection between emerging flux with pre-existing flux \citep{Chen2000,FeyM:1995,chifor_2007}. The magnetic reconnection would intern change the force balance by reducing the pressure at the reconnection location. Due to this reduced pressure the inflow structure, which is seen as one of the legs of the CME, starts to move southwards towards the reconnection region and eventually also reconnects and disappears. Later on a jet like structure emanates from the location where the inflow structure reconnects. The jet like structure is clearly 
observed in 304~{\AA} and 1600~{\AA} images. The detection of the HXR source, the southward movement of the inflow structure and rise of the CME structure in the corona is closely related to each other. 

In the beginning of the event,  the CME structure appears as blob-like, which later revealed a helical structure with a overlying structure. Therefore, the current observation also signifies that the formation of the flux rope that is likely to be a consequence of the reconnection. Moreover, for the first time we observe signature of magnetic reconnection in lines like 304~{\AA} and 1600~{\AA}. Such observations may have potential to provide information on reconnection processes occurring in partially ionised plasmas. 

\acknowledgments We thank the referee for the comments. SM and DT acknowledges the support from DST under the 
Fast Track Scheme (SERB/F/3369/2012/2013). The AIA data are courtesy of SDO (NASA) and the AIA consortium. HI is supported 
by the Grant-in-Aid for Young Scientists (B, 22740121). RHESSI work is supported by NASA contract NAS 5-98033.


\end{document}